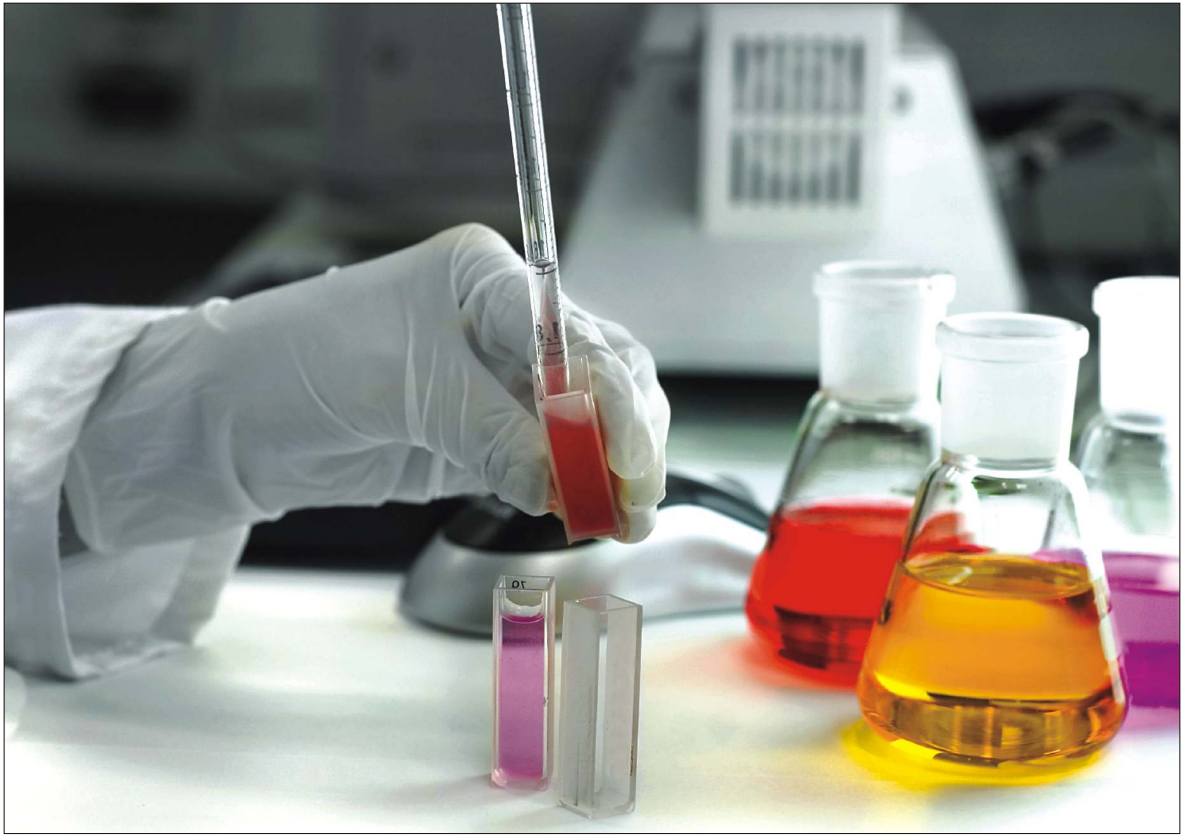

# Journal of Applied Sciences

# Research Article
# Data Mining Techniques in Predicting Breast Cancer

Hamza Saad and Nagendra Nagarur

Department of System Sciences and Industrial Engineering, The State University of New York, Binghamton, New York, USA

## Abstract

**Background and Objective:** Breast cancer, which accounts for 23% of all cancers, is threatening the communities of developing countries because of poor awareness and treatment. Early diagnosis helps a lot in the treatment of the disease. The present study conducted in order to improve the prediction process and extract the main causes impacted the breast cancer. **Materials and Methods:** Data were collected based on eight attributes for 130 Libyan women in the clinical stages infected with this disease. Data mining was used by applying six algorithms to predict disease based on clinical stages. All the algorithms gain high accuracy, but the decision tree provides the highest accuracy-diagram of decision tree utilized to build rules from each leafnode. Ranking variables applied to extract significant variables and support final rules to predict disease. **Results:** All applied algorithms were gained a high prediction with different accuracies. Rules 1, 3, 4, 5 and 9 provided a pure subset to be confirmed as significant rules. Only five input variables contributed to building rules, but not all variables have a significant impact. **Conclusion:** Tumor size plays a vital role in constructing all rules with a significant impact. Variables of inheritance, breast side and menopausal status have an insignificant impact in analysis, but they may consider remarkable findings using a different strategy of data analysis.

**INTRODUCTION**

Breast cancer is a common health problem that attacks women in the world, it is one of the most known malignancies with 23% of all types of cancers, with over one million new cases detected per year[1-2]. Roughly 4.4 million women are living with breast cancer and more than 400,000 died annually from the disease. This disease recorded 14% of all cancer deaths[3]. It is the most common cause of female death in industrialized countries[4], the second most common cause in the world and the third most common in developing countries[5]. The protection from disease is by getting an early physical exam which makes therapy more beneficial. Despite development in the strategies for disease treatment, advanced breast cancer remains incurable and the goals of therapy range from symptom palliation to extending survival.

Breast cancer is the uncontrolled development of cells in the breast. It mostly, the disease affects females, but males also suffer from the disease. Different factors can indicate the occurrence of disease. The most important factor associated with breast cancer is a family history (Inheritance). Other risk factors that can lead to the occurrence of breast cancer are food, environment demographics, marital status, health condition, breast feeding, menarche, menopause, age and a number of children. The ratio of breast cancer in different areas differs based on specific factors. Similarly, mortality rates are decreased and increased in different regions; in industrialized countries, the mortality rate is lower than in developing countries[6].

Small tumors are more be treated successfully by early detection[7-8]. Delayed detection of breast cancer is correlated with danger clinical stages and low survival percentage[9-11]. Reports in developed countries indicated that the median time to the consultation was 14-61 days[12-14]. A delay for more than three months before physician checking happened in 14-53% of cases[15]. Minority ethnicity status low socio-economic and younger age were correlated with a longer duration of symptoms[16]. Diagnosis delay was also correlated with the older age, tiny symptoms and fear to inform anyone.

In developing countries, the management of breast cancer faces social, significant medical and economic problems. The patients with breast cancer usually present with the dangers of clinical-stage, dominant presence in premenopausal status, young age, have early disease recurrence and are associated with high mortality[4]. Despite advances in the treatment, the mortality average is still high. Therefore, it is necessary to secure good cancer control by applying different strategies such as; improve understanding of early detection and find the prognostic variables, which applied with traditional factors that can predict the output of the individual patient and allow selection of appropriate therapy[16].

Libya as an example of developing countries, the statistics are 18.8 new cases for every 100,000 women annually[17]. Most of the patients in Libya present with the danger case because they fear early detection or they had not enough knowledge about the disease[17-18]. The patients usually are younger than in Europe, in line with the pattern typical in North Africa and the Middle East[18,19]. To improve the care service of breast cancer, it must get abetter understanding of the predicting causes, factors, attributes and treatment delay[20].

In recent years, patients of breast cancer have increased dramatically in developing countries for many reasons, such as; increased pollution and poor awareness of the seriousness of this disease. Many women, especially in Libya fear entering early detection, which is necessary in treating the disease. Most of the cases who admitted to the Libyan hospital are in the dangerous health situation that respond slowly to the treatment and cancer medicine. Also, the treatment of cancer in Libya is still weak and does not perfectly respond to patient's bodies that have been exhausted from the disease, leading to the patient's death. Some women in developing countries who attacked from breast cancer are survived with one breast for a limited life.

This study analyzed the disease detailing some variables based on the collected data in order to get more understanding of the main causes and analyze the breast cancer data to reach a logic decision to help new cases based on the final prediction. The study presents an integrated strategy using data mining as the main step to predict and extract important relationships from supervised attributes of breast cancer.

**MATERIALS AND METHODS**

**Study area:** The study was carried out between January-November, 2019 at the Watson School, Department of System Sciences and Industrial Engineering, State University of New York at Binghamton.

**Data mining:** Data mining and artificial intelligence have become extremely applying in medicine to help physicians extract relevant information in decision-making, especially about critical cases that depend on the main causes, which may understand by data mining (Fig. 1).

Many studies have used statistics as a traditional method of capturing and understanding data by focusing on p-value[3]. Moreover, early diagnosis is covered mainly by several papers and it is proven that early diagnosis of any disease, even cancer may increase the likelihood of treatment. The study focused on the application of data mining in predicting breast cancer using clinical stages, where it was used KNN, Tree, SVM, Neural Network, Naïve Bayes and Logistic Regression to achieve this goal. Although, all the algorithms produced high accuracy, the decision tree was chosen to complete and extend the solution. Any accuracy produced by an algorithm is due to the influence of the input variables. To understand these variables, the decision tree was exploited by drawing it and following each split to reach the decision leaf and build the appropriate rule. Based on the number of decision leaves, the rules were constructed and only the decision leaf containing its pure subset (100% classification from the same clinical stage) was confirmed as strong rule. It may not be enough to know only the essential variables that built the decision tree, but to understand each variable, feature selection was used to distinguish influential variables from weak variables and support the final solution.

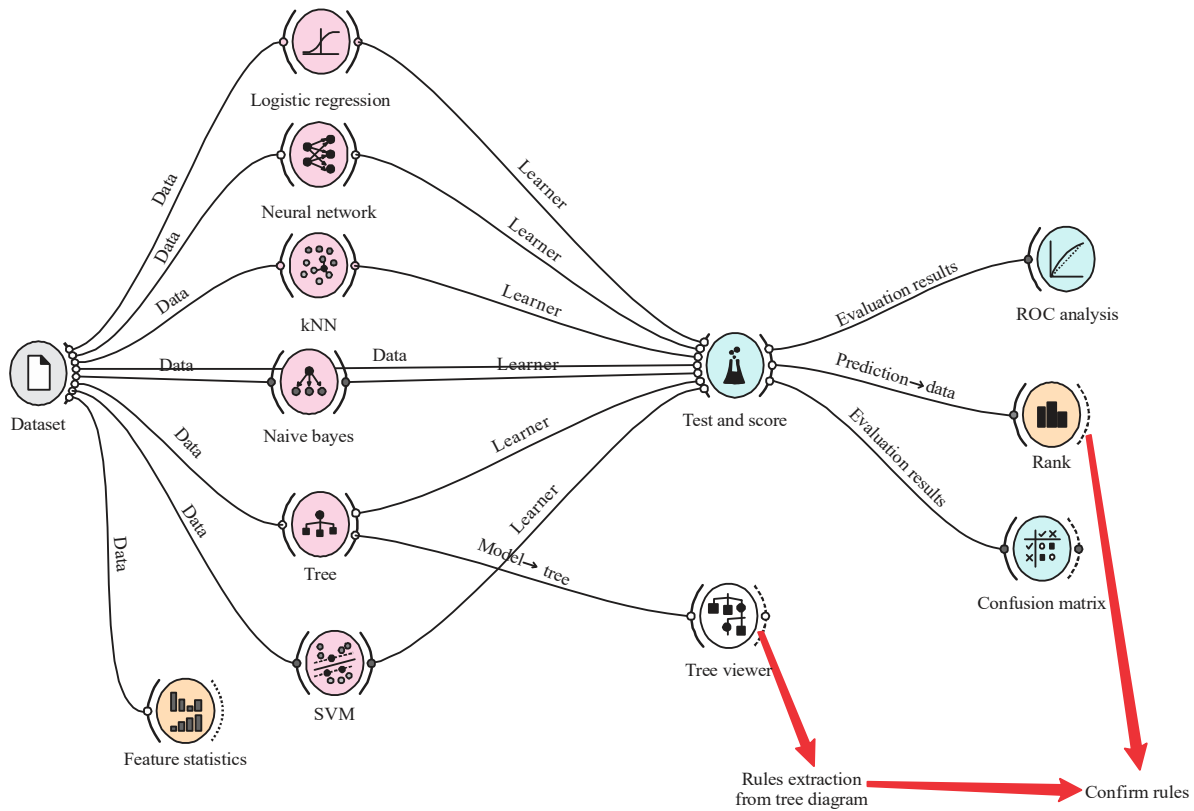

Fig. 1: Framework of data analysis by using data mining

**Software:** Orange is a component-based visual programming package for data mining, data visualization, machine learning and data analysis. Orange components called widget and they range from preprocessing, subset selection and simple data visualization to an empirical evaluation of predictive modeling and learning algorithms.

Visual programming is conducted through an interface in which workflows are established by liking user-designed widgets or predefined, while advanced users can use the orange package as a Python library for data manipulation and widget alteration.

**Summary to applied algorithms**

**KNN:** It is a simple algorithm that stored all available instances and predicts the numerical target based on the similarity measuring (distance functions). K nearest neighbors has already been applied in pattern recognition and statistical estimation at the beginning of the 1970s as non-parametric techniques.

**Tree:** The decision tree establishes models of classification or regression in the form of a tree diagram. It splits down a dataset into smaller and smaller subsets, while at the same time an associated decision tree is gradually developed. The outcome is a tree with leaf nodes and decision nodes. A decision node (e.g., tumor size) has 2 or more branches (e.g., tumor size and inheritance), each one is represented values for the tested attribute. Leaf node (e.g., clinical stages) represents the ultimate decision on the categorical target. The top most decision node in the tree that corresponds to the best predictor or input variable called the root node. Decision trees can handle and analyze numerical and categorical data.

A decision tree applied in the study to classify data based on the clinical stages according to its accuracy. The diagram of the decision tree will be utilized to build the rules based on the pure subset (100% at a leaf node). Orange software successfully applied to analyze data and generate a tree graph because it is the best software that draws the tree graph with full details and simple splitting.

**SVM:** A Support Vector Machine (SVM) conducts a classification by getting the hyperplane that maximizes the margin between the 2 instances. The cases (vectors) that defined the hyperplane called the support vectors.

**Neural network:** An Artificial Neural Network (ANN) is a system based on biological neural networks, like a brain. An ANN is comprised of an artificial neuron network (known as "nodes"). These nodes connected in network shape and strength of the connections to another is assigned in the value based on strength; the inhibition (maximum being -1 and 0) or the excitation (maximum being +1 and 0). Within each design of the node, the transfer function is calculated. Three types of neurons in an artificial neural network are input node, hidden node and the output node. The input nodes take in the information, in the form of which can be explained numerically. The information presented the activation values, where each node gives a number, the higher the number means the huge activation. This information is then passed throughout the whole network. Based on the connection weights (strengths), transfer functions and excitation or inhibition, the activation value is passed through the node to node. Each of the nodes sums the activation values it receives; it is then modifying the value based on its transfer function. The activation flowed through the network, through the hidden layer, until it reached the output nodes. Then, the output nodes reflect the input in a meaningful way to an outside world.

**Naïve Bayes:** The classifier of Naive Bayesian is generated from the "Bayes Theorem" with the independence assumptions between variables (Predictors). A Naive Bayesian classifier is easy to build, with no problematic iterative parameter estimation, which makes it is useful for substantial medicine datasets. Regardless of its simplicity, the algorithm often does surprisingly well and is widely applied because it often outperforms perfect classification methods. It can predict only a categorical output.

**Logistic regression:** The LR predicts the probability of the outcome that can include only two values (a dichotomy). The prediction is based on the application of one or more predictors (categorical and numerical). A linear regression is not fit to predict the value of a binary variable because a linear regression will predict values out the acceptable range (outside the range between 0-1), moreover, since the dichotomous experiments can include only one of two possible values for each test, the residuals will be not normally distributed about a predicted line[20-21].

**Predictor screening:** For big data and large number of variables, the application of algorithm of data mining becomes difficult, for example, neural networks become impossible to manage when the number of input variables into the model exceeded a few hundred or even less. Therefore, it is easy a practical necessity to choose and screen essential variables from among a big set of predictor variables that are most likely the utility to predict the outputs of the interest. The objective behind the Predictor Screening module is to choose a set of predictor variables based on the dependent variable from an extensive list of candidates allowing them to focus on a more professional set for further analysis. The Predictor Screening module optimally handles categorical and continuous predictors, then estimate their predictive power that can improve accuracy and get a sophisticated output using influential predictors. This study is included few variables, but the rules will build based on some variables, whether weak or have a significant impact. Predictor screening will be utilized to support rules by ranking the significant variables from the dataset[20,21].

**Data collection:** Data includes 130 patients diagnosed with one of the danger clinical stages of breast cancer. It collected between 2017 and 2018 from the same place (Oncology Hospital, Tripoli, Libya) based on the psychology and physical aspects. Eight input variables will be predicted according to the clinical stages.

Age is a numerical variable between 25 and 66 years old with mean 44.1 and standard deviation 10.325. Menopausal status divided into two distinct, perimenopause 84 patients with a percentage of 64.61% and post-menopausal 46 patients with a percentage of 35.38%. Tumor size is between 1.9 and 34, the mean is 13.88 and a standard deviation of 7.605. The breast side is included 2 distinct, the

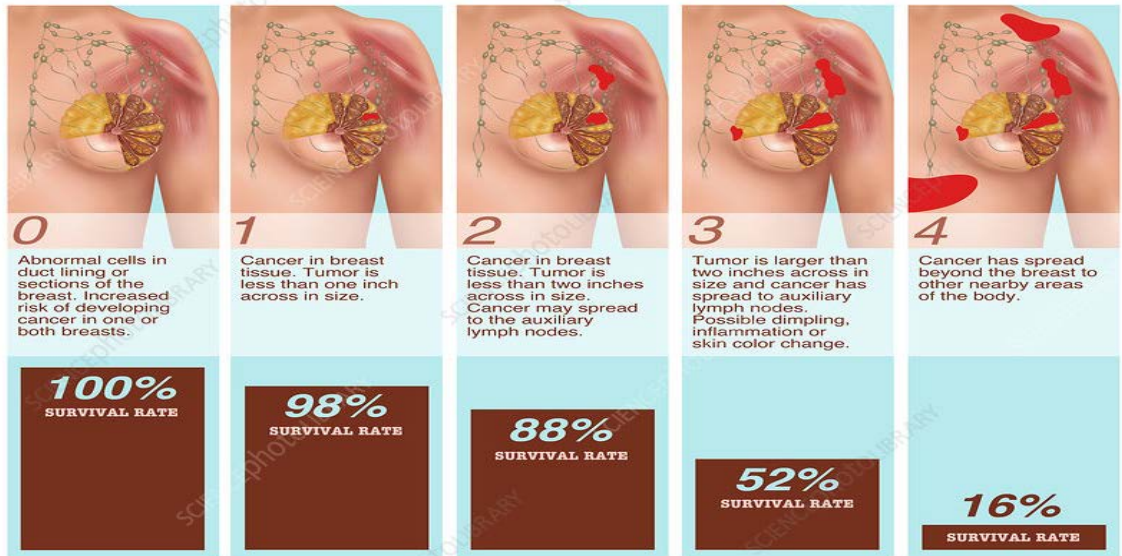

Fig. 2: Clinical stages of breast cancer
Source: Gwen Shockey/Science photo library

Table 1: Data collection

| Name | Center | Dispersion | Minimum | Maximum | Category I | Category II | Missing (%) |
|---|---|---|---|---|---|---|---|
| Age | 44.0 | 0.23 | 25.00 | 66.00 | - | - | 0 |
| Breast | Right side | 0.69 | - | - | Right, 72 | Left, 58 | 0 |
| Clinical stages | Stage 2 | 1.23 | - | - | Stage I, 6, Stage II, 44 | Stage III, 44, Stage IV, 36 | 0 |
| Early detection | No | 0.62 | - | - | Yes, 41 | No, 89 | 0 |
| Histological grade | 2.42 | 0.24 | 1.00 | 3.00 | - | - | 0 |
| Inheritance | Yes | 0.56 | - | - | Yes, 98 | No, 32 | 0 |
| Lymph node status | LN+ | 0.52 | - | - | LN+, 102 | LN-, 28 | 0 |
| Menopausal status | Perimenopause | 0.65 | - | - | Perimenopause, 84 | Postmenopausal, 46 | 0 |
| Tumor size | 13.88 | 0.55 | 1.90 | 34.00 | - | - | 0 |

LN+ and LN: Status of Lymph Node (LN) whether positive or negative

right side includes 72 patients with 55.38% and the left side includes 58 patients with 44.61%. Lymph Node (LN) status included two attributes, LN- with 28 patients and percentage 21.53%, LN+ with 102 patients and percentage 78.46%. The histological grade is a numerical variable, the grade is between 1 and 3, the mean is 2.415 and the standard deviation 0.581. Early detection, 41 women diagnosed early with no danger stage and 89 patients did not diagnose early with the disease. The inheritance includes 2 categories, Yes included 98 patients with a percentage of 75.38% and No included 32 patients with a percentage of 24.61%. Moreover, the clinical stage represents the response variable to predict all data by using data mining, it divided into four stages. Stage I with 6 patients and percentage 4.61%, stage II with 44 patients and percentage 33.84%, stage III with 44 patients and percentage 33.84% and stage IV with 36 patients with percentage 27.69%. It can be shown (from Johns Hopkins Medicine) in Fig. 2. Table 1 presents the statistics for applied data.

This study did not consider stage 0 in the data of study because all the patients in the hospital of case study were ranked from I to IV:

**Stage 0:** The cancer is only appeared inside the milk duct. This stage is non-invasive, it includes ductal carcinoma *in situ*

**Stage I:** Includes small tumors that are only affecting a small area of the sentinel lymph node

**Stage II:** Includes large tumors that are affecting some nearby lymph nodes

**Stage III:** The tumors are growing into surrounding tissues like muscle, breast skin and lymph nodes

**Stage IV:** The tumors are started in the breast and spreading to the other parts of the body (Medical News Today)

Waste time without detecting disease, the risk of death increases, especially if the patient neglected herself without getting treatment or taking ineffective medicine that is normal

Table 2: Results for different algorithms based on the response of clinical stages

| Algorithm | AUC | CA | F1 | Precision | Recall | Algorithm ranking |
|---|---|---|---|---|---|---|
| KNN | 0.926 | 0.815 | 0.797 | 0.784 | 0.815 | 4 |
| Tree | 0.949 | 0.900 | 0.900 | 0.901 | 0.900 | 1 |
| SVM | 0.950 | 0.792 | 0.772 | 0.755 | 0.792 | 6 |
| Neural network | 0.928 | 0.792 | 0.776 | 0.761 | 0.792 | 5 |
| Naïve Bayes | 0.956 | 0.823 | 0.829 | 0.849 | 0.823 | 3 |
| Logistic regression | 0.951 | 0.838 | 0.827 | 0.844 | 0.838 | 2 |

AUC: Area under the curve, CA: Classification accuracy, F1: F measure in developing countries. Therefore, early detection should be an essential requirement for all women, especially those who have relative history.

Table 3: Confusion matrix

| Stages | Stage I | Stage II | Stage III | Stage IV | Sum |
|---|---|---|---|---|---|
| I | 5 | 1 | 0 | 0 | 6 |
| II | 0 | 37 | 7 | 0 | 44 |
| III | 0 | 5 | 39 | 0 | 44 |
| IV | 0 | 0 | 0 | 36 | 36 |
| Sum | 5 | 43 | 46 | 36 | 130 |

Data were collected based on the variable of clinical stages that is used as an output to predict disease using eight predictors to classify the next patients with the correct clinical stage. The clinical stages variable was tested before confirming the final solution to extract high accuracy from the dataset.

The performances of the algorithms ranged from high to low as follows; Tree, Logistic Regression, Naïve Bayes, KNN, Neural Network and SVM that gives the lowest performance.

Some equations that applied by algorithms to measure and validate algorithm performance are:

$$\text{Recall} = \frac{Tp}{Tp+Fn}$$

$$\text{Precision} = \frac{Tp}{Tp+Fn}$$

$$F1 = 2*\frac{\text{Precision}*\text{recall}}{\text{Precision}+\text{recall}}$$

$$\text{True negative rate} = \frac{Tn}{Tn+Fp}$$

where, Tp is a true positive, Fn is a false negative, Fp is false positive and Tn is a true negative rate.

**RESULTS AND DISCUSSION**

**Data mining in prediction:** Six algorithms were experimented to extract the best accuracy and figure out the characteristics of the classification using cross-validation = 10 folds. For small data available, K-folds cross-validation used to achieve an unbiased estimation of the model performance by dividing the limited data into equal sizes for K subsets. Each time leaving out one of the subsets from the training set and use it as the test set. Data of breast cancer split in the same sizes. The best accuracy was obtained from the Decision tree. The results of the algorithms are shown in Table 2.

To evaluate, analyze and validate dataset, five measures were shown in Table 2, Area Under the Curve (AUC), Classification accuracy (CA, F measure (F1) and recall)

Algorithms of data mining that applied in this study were handled data of breast cancer without preprocessing because it is real-world data, complete, consistent and no missed instances. Data of study was collected based on the clinical stages with limited scopes and the number of patients-the highest AUC (Area Under The Curve) recorded by Naïve Bayes with accuracy 0.956. The lowest AUC recorded by KNN (K-Nearest Neighbors) with accuracy 0.926. However, all algorithms provided high performance, whether in handling or analyzing data to generate high accuracy by each one. In CA (Classification Accuracy), tree algorithm generated the highest accuracy with 90% and the lowest accuracy of CA was recorded by 2 algorithms SVM (Support Vector Machine) and NN (Neural Network). So, in Table 2, at precision indicator, the tree algorithm was confirmed as the best algorithm by giving accuracy 90%.

All applied algorithms provided prediction with different performances.

In the confusion matrix of decision tree Table 3, the output variable includes four clinical stages and each stage classified with a particular accuracy value, the accuracy of each stage will calculate to support reliable results. The confusion matrix is presented as actual and predicted values; according to these values, the accuracy of each stage has been calculated.

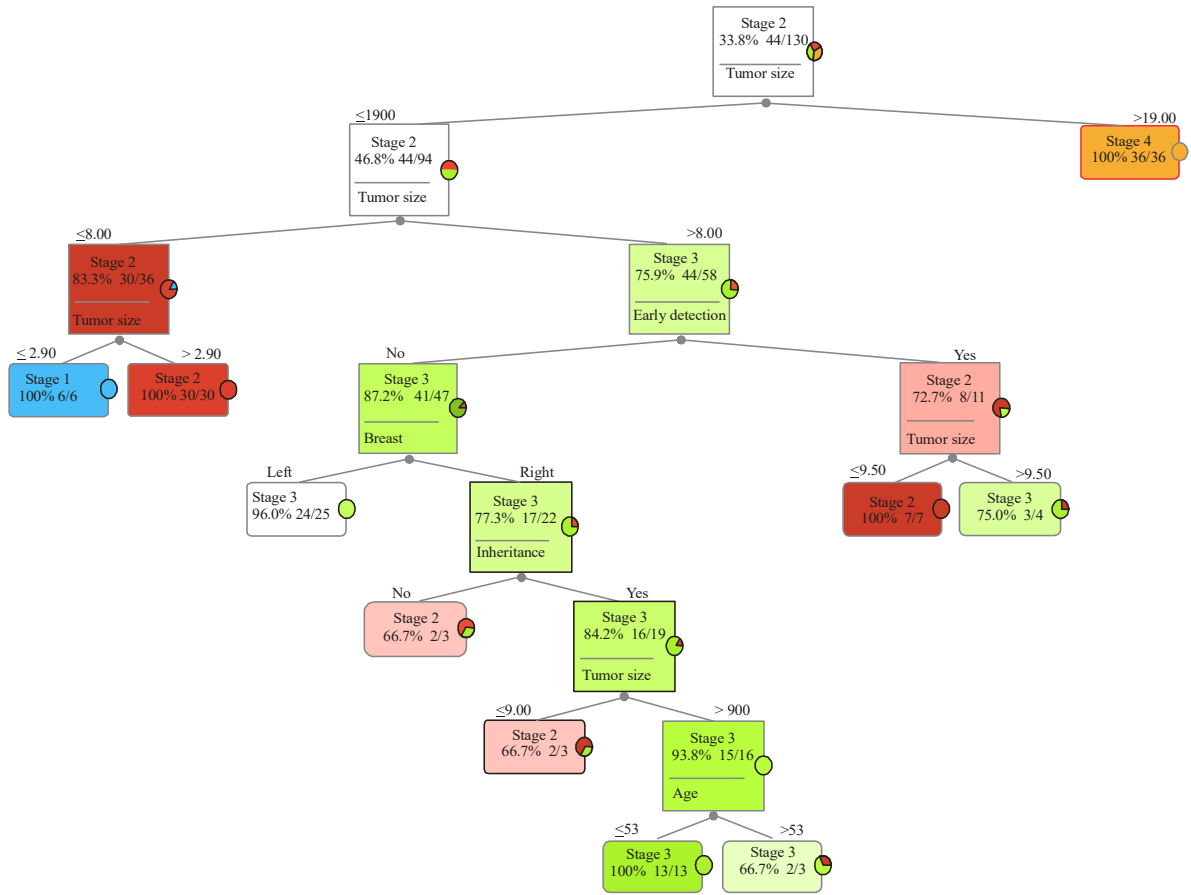

Fig. 3: Tree graph from decision tree

Each stage gets a specific prediction in classification and according to each prediction, a vital relationship between inputs and output can be defined.

In Stage I, 5 instances classified correctly and only 1 instance is miss-classified, the overall accuracy for stage I is 83.33%. In Stage II, 37 instances classified correctly and 7 instances are miss-classified, the overall accuracy for this stage is 84.09%. For Stage III, 39 instances classified correctly and 5 instances are miss-classified, the overall accuracy for stage III is 88.63%. Moreover, Stage IV, 36 instances classified correctly and there is no miss-classified, the overall accuracy for stage IV is 100%.

Data is for 130 patients, by calculating miss-classified from each stage = 1+7+5+0 = 13 and correctly classified = 5+37+39+36 = 117. Then:

$$\text{Total accuracy} = \frac{\text{Correctly classified}}{\text{Total patient's sample}} = \frac{117}{130} = 0.9 = 90\%$$

So, the overall accuracy using the decision tree is 90%. Figure 3 showed a graph of the decision tree that will utilize to extract the rules in order to support the final decision.

The fitted predictor that can provide more splits is tumor size as shown in Fig. 3. The split is started from stage II as response and tumor size as a predictor. From the first splitting, the full category of Stage IV has predicted and ended with pure classification (36 of 36). Again, the decision tree estimated tumor size predictor to build a new split between (tumor size and early detection). Predictor of early detection extended the split to many splits. However, tumor size provided a pure subset at Stage I and Stage II with end splitting. Predictors of breast and tumor size have split from early detection to provide more splits by breast and stop splitting by tumor size. At tumor size split, pure subset at Stage II and impure classification at Stage III from the left side of the breast. On the right side, the split started again by inheritance to get impure classification at stage II. Stage II got another impure classification to start another split by age predictor. The split has ended at age predictor to get a pure classification at stage III and impure classification at stage III. (Sometimes leaf node has the same category like stage I and stage I because the split is generated based on predictors or input variables to predict the clinical stages).

Table 4: Impact of each variable in the dataset (significant variables distinguished in the shaded part)

| Input predictor | Chi-square | p-value | Variable number |
|---|---|---|---|
| Histological grade | 74.5812 | 0.000000 | 6 |
| Early detection | 68.2713 | 0.000000 | 9 |
| Tumor size | 211.9512 | 0.000000 | 3 |
| Lymph Node (LN) status | 48.8155 | 0.000000 | 5 |
| Age | 34.7038 | 0.000520 | 1 |
| Breast side | 5.8531 | 0.118980 | 4 |
| Inheritance | 3.5891 | 0.309390 | 8 |
| Menopausal status | 1.8924 | 0.595040 | 2 |

Two predictors, histological grade and lymph node status have ranked as the most significant variables in the dataset, but they are not estimated by the algorithm of decision tree to split or classify because they are not fitted with the splitting strategy. Furthermore, predictors of inheritance and breast side have ranked as insignificant or they have a weak impact on the clinical stages, so they estimated by decision tree for split and build more decision leaves. They build more split with impure classification.

**Predictor screening:** Before building rules from decision leaves or use a decision tree to build a tree structure, the impact of each variable in the data set must figure out. To know that, chi-square and p-value calculated to rank each variable according to the output (Clinical stages). Based on the results, variables of the breast side, inheritance and menopausal status did not have a significant impact on the output. However, variables of histological grade, early detection, tumor size, Lymph Node (LN) status and age provided a significant effect on the output. This rank applied to support extracted rules from the decision tree. Each rule was constructed using one of these variables and then the efficiency of each variable help to support final rules and results. Table 4 showed the rank of the effect of each variable based on chi-square and p-value.

Data mining does not end when predicting the next patient based on existing data, but its role continues to extract the significant variables that may help to support the decision. However, some variables contributed to predicting the clinical stages to varying degrees from weak to strong. By neglecting weak variables and focusing on strong variables, more robust expectations can be built by focusing on selected variables.

**Rules extraction:** From the graph of the decision tree, more knowledge can be obtained by extracting some rules and support these rules by the accuracy and effect of each variable on the final prediction:

- If tumor size >19, then clinical stage = stage IV (100% pure subset)
- If tumor size <= 19 and >8, early detection = Yes, tumor size >9.5, then clinical stage = stage III (75% pure)
- If tumor size <= 19 and >8, early detection =Yes, tumor size <= 9.5, then clinical stage = stage II (100% pure subset)
- If tumor size <= 19 and <=8, tumor size >2.9, then clinical stage = stage II (100% pure subset)
- If tumor size < = 19 and < = 8, tumor size < = 2.9, then clinical stage = stage I (100% pure subset)
- If tumor size < = 19 and >8, early detection = No, breast = left side, then clinical stage = stage III (96% pure)
- If tumor size <= 19 and >8, early detection = No, breast side = right, inheritance = Yes, tumor size >9, age>53, then clinical stage = stage III (66.7% pure)
- If tumor size< = 19 and >8, early detection = No, breast side = right, inheritance = No, then clinical stage = stage II (66.7% pure)
- If tumor size < = 19 and >8, early detection = No, breast = right side, inheritance = Yes, tumor size >9, age < = 53, then clinical stage = stage III (100% pure subset)
- If tumor size < = 19 and >8, early detection = No, breast = right side, inheritance = Yes, tumor size>9, age>53, then clinical stage = stage III (66.7% pure)

Few predictors estimated by algorithm to generate 90% accuracy. Rules extraction and predictor screening have joined in Table 5 to measure how the significant predictors impacted on the clinical stages.

Rules 1, 3, 4, 5 and 9 have a 100% pure subset for each category of the clinical stage. However, the final accuracy of the algorithm does not provide accuracy more than 90% because it affected by all input variables whether these variables are significant or insignificant (weak predictors increased the split in the tree with impure subsets). According to rule 1, there is only one variable effect in stage IV, but tumor size must exceed 19. Hence, patients with tumor size more than 19 can be predicted to enter one of the danger stages of breast cancer. For Rule 3, there are 2 predictors effect in stage II like tumor size and early detection, but tumor size has a higher impact and the patient must get an early diagnosis. For

Table 5: Effect of each variable in rules construction

| Variables | Rules | | | | |
|---|---|---|---|---|---|
| | 1 | 3 | 4 | 5 | 9 |
| Tumor size | Significant | Significant | Significant | Significant | Significant |
| Early detection | | Significant | | | Significant |
| Breast side | | | | | Insignificant |
| Age | | | | | Significant |
| Inheritance | | | | | Insignificant |
| Prediction→ | Stage IV | Stage II | Stage II | Stage I | Stage III |

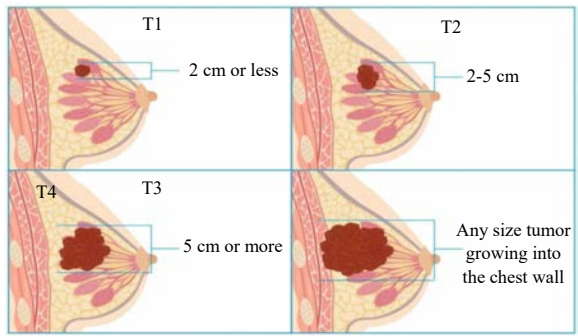

Fig. 4: Tumor size for each stage
Source: Medical news today

Rule 4 and Rule 5, only tumor size impacted on stage II and stage I, whether tumor size is between 8 and more significant than 2.9 for stage II or between 8 and less than 2.9 for stage I. However, in Rule 9, there are many variables impact on final classification of stage III such as; tumor size must be greater than 8, early detection with late diagnosis, right breast side and there are close blood relatives in the same family with the disease (inheritance). Figure 4 presented how does tumor damage the breast if the disease does not detect early.

## CONCLUSION

Breast cancer is a common cancer for females in developing countries. Cancer progresses to the dangerous stage with the time, in which the tumor spreads to the body. Early detection leads to control of the disease, but neglect it means an increased chance of death. Many factors are related to the clinical stages, but the tumor size factor is a major problem that requires the patient greater care for checking up each time (between 3 months to a year). In developing countries, tumor size is a big problem threatening patients who are less educated to deal with cancer diseases. The accuracy is high from each algorithm because data is strongly related to the output (clinical stages).

## SIGNIFICANCE STATEMENT

There are more factors that may affect the disease. Family history factor is the big challenge to the people for considering an early detection. The tumor size is increased from 0 to IV in case of neglecting early detection or using ineffective medicine due to the high price of the cancer medicine. There are significant and direct relationships between the tumor size and other factors. Tumor size can play a main role to predict a specific clinical stage.